%% file: main.tex
\documentclass[sigconf]{acmart}

\setcopyright{none}
\renewcommand\footnotetextcopyrightpermission[1]{}
\AtBeginDocument{%
  }

\usepackage{tabularx}
\usepackage{booktabs}
\usepackage[table]{xcolor}
\definecolor{lightgreen}{HTML}{E6F4EA}
\usepackage{tikz}
\usepackage{pgfplots}
\pgfplotsset{compat=1.18}

\newcommand{\method}{SAFEdit}

\begin{document}

\title{SAFEdit: Does Multi-Agent Decomposition Resolve the Reliability Challenges of Instructed Code Editing?}

\author{Noam Tarshish}
\affiliation{%
  \institution{Ben-Gurion University of the Negev}
  \city{Beer Sheva}
  \country{Israel}}
\email{noamtars@post.bgu.ac.il}

\author{Nofar Selouk}
\affiliation{%
  \institution{Ben-Gurion University of the Negev}
  \city{Beer Sheva}
  \country{Israel}}
\email{nofarse@post.bgu.ac.il}

\author{Daniel Hodisan}
\affiliation{%
  \institution{Ben-Gurion University of the Negev}
  \city{Beer Sheva}
  \country{Israel}}
\email{hodisan@post.bgu.ac.il}

\author{Bar Ezra Gafniel}
\affiliation{%
  \institution{Ben-Gurion University of the Negev}
  \city{Beer Sheva}
  \country{Israel}}
\email{ezrabar@post.bgu.ac.il}

\author{Yuval Elovici}
\affiliation{%
  \institution{Ben-Gurion University of the Negev}
  \city{Beer Sheva}
  \country{Israel}}
\email{elovici@post.bgu.ac.il}

\author{Asaf Shabtai}
\affiliation{%
  \institution{Ben-Gurion University of the Negev}
  \city{Beer Sheva}
  \country{Israel}}
\email{shabtaia@bgu.ac.il}

\author{Eliya Nachmani}
\affiliation{%
  \institution{Ben-Gurion University of the Negev}
  \city{Beer Sheva}
  \country{Israel}}
\email{eliyanac@bgu.ac.il}

\input{sections/0_abstract/abstract}

\maketitle
\thispagestyle{empty}
\pagestyle{empty}

\vspace{-1em}
\input{sections/1_introduction/introduction}

\input{sections/2_related_work/related_work}

\input{sections/3_methodology/methodology}

\input{sections/4_evaluation/evaluation}

\input{sections/5_results/results}
\input{sections/6_discussion/discussion}
\input{sections/7_conclusion/conclusion}

\bibliographystyle{ACM-Reference-Format}
\bibliography{references}

\end{document}

%% file: sections/0_abstract/abstract.tex
\begin{abstract}
Instructed code editing is a significant challenge for large language models (LLMs). 
On the EditBench benchmark, 39 of 40 evaluated models obtain a task success rate (TSR) below 60\%, highlighting a gap between general code generation and the ability to perform instruction-driven editing under executable test constraints. 
To address this, we propose \textbf{SAFEdit}, a multi-agent framework for instructed code editing that decomposes the editing process into specialized roles to improve reliability and reduce unintended code changes. 
A \textit{Planner} Agent produces an explicit, visibility-aware edit plan, an \textit{Editor} Agent applies minimal, literal code modifications, and a \textit{Verifier} Agent executes real test runs. 
When tests fail, SAFEdit uses a Failure Abstraction Layer (FAL) as part of the framework to transform raw test logs into structured diagnostic feedback, which is fed back to the Editor to support iterative refinement.
We compare SAFEdit against both prior single-model results reported for EditBench and an implemented ReAct single-agent baseline under the same evaluation conditions.
We used EditBench to evaluate SAFEdit on 445 code editing instances in five languages (English, Polish, Spanish, Chinese, and Russian) under varying spatial context variants. 
SAFEdit achieved \textbf{68.6\%} TSR, outperforming the single-model baseline by \textbf{+3.8} percentage points and the ReAct single-agent baseline by \textbf{+8.6} percentage points. 
The iterative refinement loop was found to contribute \textbf{+17.4} percentage points to SAFEdit's overall success rate. 
SAFEdit's automated error analysis further indicates a reduction in instruction-level hallucinations compared to single-agent approaches, providing an additional framework component for interpreting failures beyond pass/fail outcomes.
\end{abstract}

%% file: sections/1_introduction/introduction.tex
\section{Introduction}
\label{sec:intro}
Software developers spend a large amount of time making targeted modifications to existing code: fixing bugs, adding parameters, refactoring functions, and extending classes to support new behaviors. 
As a result, automated assistance for editing existing code has become a practical and high-impact LLM application\cite{chi2025editbenchevaluatingllmabilities,guo2024codeeditorbench}. 
Unlike generative coding tasks, instructed code editing requires precise reasoning over existing structure, preservation of invariants, and minimal, targeted modifications that satisfy explicit correctness constraints\cite{chi2025editbenchevaluatingllmabilities,guo2024codeeditorbench,chen2021evaluatinglargelanguagemodels}.

Recent evaluations of LLMs point to a concerning reality: even cutting-edge models struggle to reliably perform instructed code editing.
On the EditBench benchmark~\cite{chi2025editbenchevaluatingllmabilities}, a collection of 540 real-world editing tasks drawn from developer Integrated Development Environment (IDE) sessions and validated by executable test suites, 39 of 40 evaluated models were found to obtain a task success rate (TSR) of less than a 60\%. 
The best single-model baseline (claude-sonnet-4) achieved a TSR of just 64.8\% under the most informative context variant. 
These results indicate that model capability alone is insufficient. 

Trustworthy code editing requires structured reasoning about the intended modification, careful restriction of changes to the relevant regions, and validation through actual test execution.

Recent studies have demonstrated the ability of multi-agent architectures to perform complex coding tasks by decomposing them into specialized sub-tasks~\cite{huang2024agentcodermultiagentbasedcodegeneration,chen2024coder}. 
However, existing approaches focus primarily on code generation or full-issue resolution~\cite{jimenez2024swebenchlanguagemodelsresolve}, with no methods specifically designed for or evaluated on the structured, context-sensitive, instructed editing setting.

This paper presents SAFEdit (\textbf{S}tructured \textbf{A}gentic \textbf{F}ramework for Trustworthy Code \textbf{Edit}ing), a multi-agent system that decomposes instructed code editing into three sub-tasks that are executed by specialized agents orchestrated via the CrewAI framework~\cite{barbarroxa2024benchmarking}: a \textbf{Planner} that translates the instruction into a structured, visibility-aware edit plan without generating any code, an \textbf{Editor} that performs controlled modification by applying the plan minimally and literally, and a \textbf{Verifier} that performs execution-based validation via real test runs in a sandboxed environment. 
When tests fail (not all of the unit tests passed), a failure abstraction layer (FAL) transforms raw test logs into structured diagnostic feedback that identifies the failed test, exception type, expected vs.\ actual values, and a suggested repair action, which is passed back to the Editor; a maximum of three refinement iterations are permitted. 

We evaluated SAFEdit on 445 code editing instances in five languages (English, Polish, Spanish, Chinese, and Russian), using the same EditBench benchmark visibility variants and test infrastructure to ensure comparability with the original EditBench benchmark.
SAFEdit achieves an average TSR of \textbf{68.6\%} across all five languages, surpassing the ReAct single-agent baseline by \textbf{+8.6pp} on average and the best single-model baseline (claude-sonnet-4, 64.8\%) by \textbf{+3.8pp}. 

Our main contributions are summarized as follows:
\begin{itemize}

\item \textbf{Architectural Decomposition for Instructed Code Editing.}
We introduce SAFEdit, a structured agentic framework that divides the instructed code editing task into three sub-tasks: planning, editing, and verification, which are performed by three specialized agents. 
Our results show that this architectural structure yields consistent TSR improvements over strong single-agent and LLM baselines while holding the backbone model constant, suggesting that structured decomposition, together with execution-grounded iterative refinement, contributes meaningfully to performance beyond backbone choice alone.

\item \textbf{Robustness to Context Exposure.}
Across all evaluated languages, both SAFEdit and ReAct exhibit a systematic decline in the TSR as context exposure is reduced, consistent with the intuition that less contextual information makes the task harder. This contrasts with the original EditBench findings, where context-level effects were model-dependent and often inconsistent, suggesting that our multi-language evaluation reveals a more stable and interpretable signal.

\item \textbf{Iterative Verification as a Structural Mechanism.}
We demonstrate that verification-driven refinement is not merely corrective but structurally transformative, contributing on average +17.4 percentage points over first-pass performance and eliminating regression errors observed in the centralized reasoning baselines.

\item \textbf{Failure Redistribution Through Role Separation.}
Using an error taxonomy, we show that SAFEdit reshapes the distribution of failure categories: reducing regression errors entirely and shifting failures from instruction\allowbreak-level hallucination toward implementation-level refinement gaps, revealing qualitative differences in reasoning behavior beyond aggregate success rates.

\end{itemize}

%% file: sections/2_related_work/related_work.tex
\section{Related Work}

\subsection{Instructed Code Editing}

The instructed code editing task is fundamentally distinct from traditional code generation: models must interpret natural-language instructions to modify \emph{existing} code, rather than synthesize new programs from scratch~\cite{cassano2024editevaluatingabilitylarge}.
In the traditional setting, widely used \emph{evaluations} of code-focused LLMs, including the Codex benchmark suite, assess a model's ability to perform function-level generation, while subsequent program-synthesis studies often assessed performance in isolated settings without prior editing history~\cite{chen2021evaluatinglargelanguagemodels,austin2021programsynthesislargelanguage}.  
This generation-centric evaluation paradigm does not reflect how developers interact with code assistants in practice, where the dominant workflow is to iteratively edit and refine existing files in response to natural-language instructions. 
More recent benchmarks such as SWE-bench evaluate whether models can resolve real GitHub issues end-to-end; doing so requires repository-level reasoning, including navigating multiple files and using large code contexts during inference~\cite{jimenez2024swebenchlanguagemodelsresolve}. 
While this setting is realistic and challenging, it targets full issue resolution at the repository level rather than localized edits to a given snippet under a concrete instruction. 
In contrast, instructed code editing focuses on such localized, instruction-driven changes within the context of an existing file, which is precisely the scenario captured by EditBench and our proposed SAFEdit framework~\cite{chi2025editbenchevaluatingllmabilities}.

In addition to existing general code-generation benchmarks, several datasets were designed specifically to evaluate instructed code editing. 
Can It Edit? serves as a benchmark dataset and evaluation protocol for measuring how well LLMs follow code-editing instructions: given an input program and a natural-language edit description, models must produce an updated program that satisfies the requested change~\cite{cassano2024editevaluatingabilitylarge}. 
The accompanying study analyzed model performance on this benchmark and found that LLMs often fail to apply the requested edits precisely, indicating that instruction adherence in editing scenarios remains a significant open challenge~\cite{cassano2024editevaluatingabilitylarge}. 
CodeEditorBench, a benchmark suite of editing sub-tasks (e.g., debugging, polishing, refactoring) covering multiple languages and models, is used to assess models' abilities to perform a wide range of code-editing capabilities~\cite{guo2024codeeditorbench}. However, both Can It Edit? and CodeEditorBench were built largely from synthetic or semi-artificial task constructions and therefore do not fully capture the ambiguity, noise, and heterogeneity of real developer instructions issued in integrated development environments.

EditBench is a benchmark dataset of 540 real-world instructed code-editing problems assembled from nearly 500 developers via a VS Code extension, spanning Python and JavaScript and covering categories such as bug fixing, feature addition, and optimization~\cite{chi2025editbenchevaluatingllmabilities}. The dataset includes three visibility variants (CODE ONLY, HIGHLIGHT, HIGHLIGHT + CURSOR) that control whether the model sees the entire file, a highlighted region, or a specific cursor location, thereby probing both spatial grounding and use of contextual information~\cite{chi2025editbenchevaluatingllmabilities}. Evaluation is performed using a containerized test environment, yielding automated pass/fail outcomes that reflect common testing practices in software development~\cite{chi2025editbenchevaluatingllmabilities}. We start from the full EditBench dataset and evaluate on a curated subset of 445 tasks after uniform filtering. The SAFEdit framework's Planner agent is designed to operate under the three visibility variants, and the framework relies on test-based verification as the primary indicator of editing correctness.

\subsection{Iterative Program Repair and Execution-Driven Refinement}

A growing body of research has studied iterative program repair with LLMs, where models repeatedly generate patches, receive execution feedback, and refine their outputs.
\emph{The Art of Repair: Optimizing Iterative Program Repair with Instruction-Tuned Models}~\cite{ruiz2025art} showed that instruction-tuned models can benefit substantially from test-based feedback across multiple repair rounds, but also that there is a point of diminishing returns: beyond a certain number of iterations, additional rounds increase computational cost while yielding smaller gains and, in some settings, even reduce effectiveness due to overfitting. 
Other studies combined LLMs with program analysis and test generation; Jiang et al.\ proposed LeDex, which includes a training framework that improves models' self-debugging and explanation abilities via execution-verified refinement trajectories~\cite{jiang2025ledextrainingllmsbetter}. 
Prasad et al.\ generated unit tests to better support automated debugging~\cite{prasad2025learninggenerateunittests}. 
Gu et al.\ investigated an iterative hybrid program-analysis approach for LLM-based test generation, alternating between analysis and generation steps in a feedback loop to improve coverage over successive iterations~\cite{gu2025llmtestgenerationiterative}.

These studies guide key aspects of SAFEdit's design. 
SAFEdit includes a test-grounded Verifier that executes real test suites, ensuring that refinement is driven by actual execution outcomes rather than heuristic signals. 
In addition, SAFEdit's refinement loop is capped at three iterations, reflecting prior observations that a modest number of iterations often provides a good trade-off between accuracy and computational cost in iterative repair and test-generation pipelines~\cite{ruiz2025art,jiang2025ledextrainingllmsbetter,prasad2025learninggenerateunittests}.

\subsection{Multi-Agent Architectures for Code Tasks}

Multi-agent LLM systems have been proposed to decompose complex software tasks into specialized agents that are coordinated via tool calls and message passing. 
AgentCoder is a multi-agent code generation system that includes programmer, test-designer, and executor agents that iteratively generate code, design tests, and run them, with the executor passing error information back to the programmer to refine subsequent revisions~\cite{huang2024agentcodermultiagentbasedcodegeneration}.
CodeR similarly structures issue resolution on SWE-bench repositories, using task graphs and multiple agents responsible for understanding, planning, patching, and verification~\cite{chen2024coder}.

The orchestration of such specialized agents is a central challenge in multi-agent systems.
A benchmark study by Barbarroxa et al.\ evaluated orchestration frameworks such as AutoGen, CrewAI, and TaskWeaver, reporting that they are well suited to coordinate specialized LLM agents in domains such as software engineering~\cite{barbarroxa2024benchmarking}.  

SAFEdit is situated within this multi-agent landscape: it decomposes instructed code editing into Planner, Editor, and Verifier agents orchestrated via CrewAI, where the Planner produces structured, visibility-aware edit plans, the Editor applies minimal edits constrained by the plan, and the Verifier executes tests to provide ground-truth feedback. 
We adopt CrewAI in SAFEdit because it provides a lightweight, role-centric abstraction for defining agents and their communication patterns, and it integrates easily with the tool-calling and orchestration primitives needed for our code-editing pipeline~\cite{barbarroxa2024benchmarking}. SAFEdit's structured multi-agent design enables clearer responsibility boundaries and more controllable editing workflows compared to unstructured single-agent approaches~\cite{barbarroxa2024benchmarking}.

%% file: sections/3_methodology/methodology.tex
\section{Methodology}
\label{sec:methodology}

In this section, we describe the design and implementation of \textbf{SAFEdit}, our proposed multi-agent framework for instructed code editing. 
Its editing pipeline, which is organized into specialized agents with distinct roles, includes structured feedback mechanisms and automated failure analysis to improve both accuracy and explainability.

\subsection{Task Definition and Input Representation}
\label{sec:problem}

SAFEdit addresses the task of instructed code editing, which is defined as follows: 
given a natural-language instruction, the original source code, and an associated unit test suite, generate a modified version of the code that satisfies all tests.

In instructed code editing, the amount of spatial context provided to the editing agent determines how precisely the required modifications can be localized. 
To evaluate the impact of context availability, we adopt both the dataset and the EditBench evaluation protocol with additional metrics~\cite{chi2025editbenchevaluatingllmabilities} and evaluate SAFEdit under three visibility variants representing different levels of edit localization:

\begin{itemize}
    \item \textbf{CODE ONLY (File-Level)}: The agent receives the entire source file without any indication of which region requires modification.
    \item \textbf{HIGHLIGHT (Function-Level)}: The relevant function or code block is marked, narrowing the search space.
    \item \textbf{HIGHLIGHT + CURSOR\_POSITION (Line-Level)}: A cursor marker pinpoints the exact line where the edit should occur, providing maximum spatial grounding.
\end{itemize}

These variants represent different amounts of contextual information available during editing and allow us to evaluate SAFEdit’s performance under varying context conditions.

\begin{figure*}[!h]
    \centering
    \includegraphics[width=\textwidth]{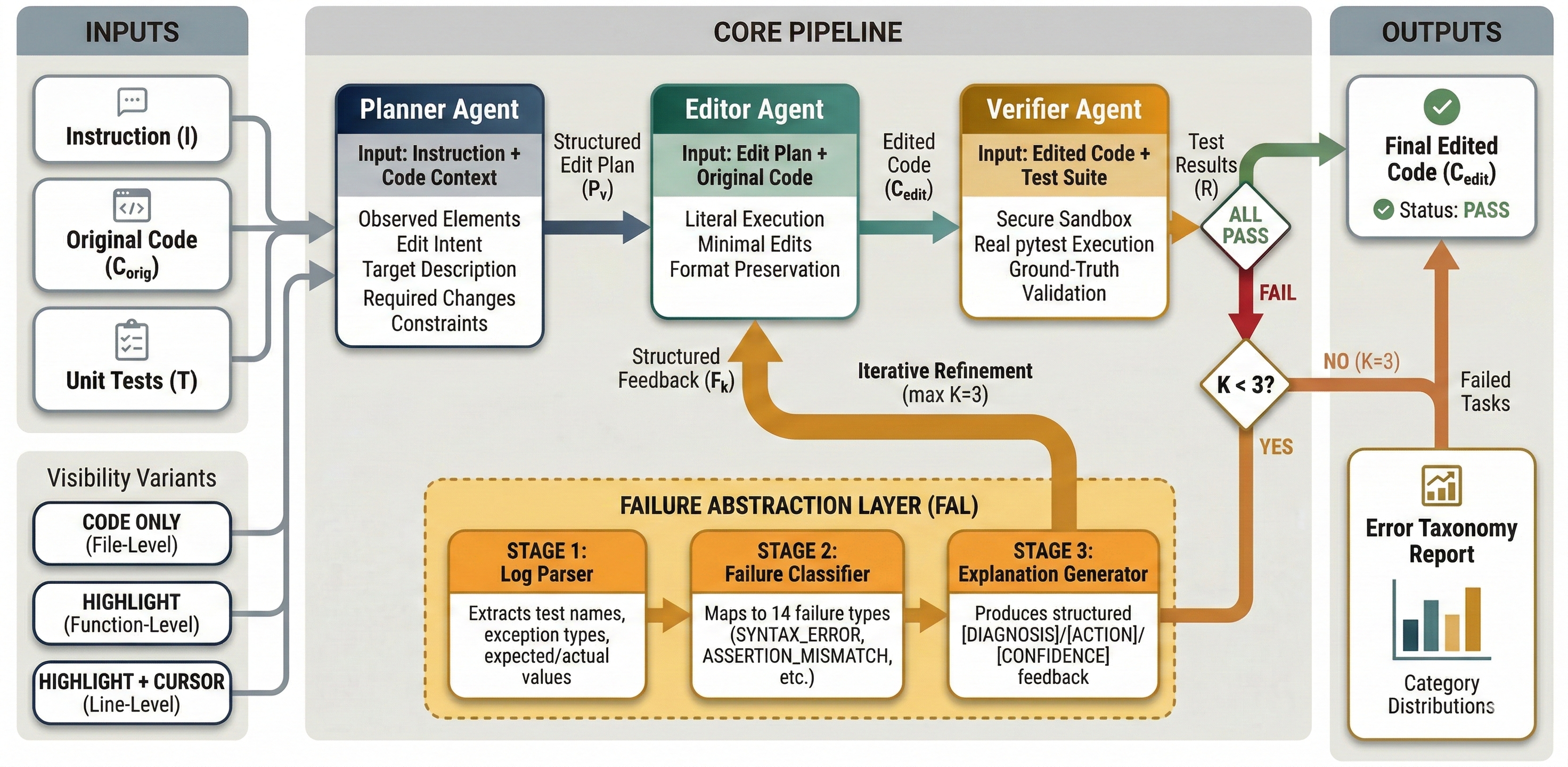}
    \caption{Overview of the SAFEdit framework. The pipeline organizes the editing task into three specialized agents (Planner, Editor, Verifier), which are connected in an iterative refinement loop. 
    The FAL transforms raw test output into structured feedback, and the error taxonomy classifies failure root causes for qualitative analysis.}
    \label{fig:methodology_overview}
\end{figure*}
\subsection{SAFEdit Design and Implementation}
\label{sec:architecture}
SAFEdit consists of three specialized agents: Planner, Editor, and Verifier, whose interaction is orchestrated using the CrewAI framework~\cite{barbarroxa2024benchmarking}. 
The agents are connected through an iterative refinement loop, in which test-based feedback is used to revise edits when necessary. 
The overall architecture is illustrated in Figure~\ref{fig:methodology_overview}.

\subsubsection{Planner Agent}
\label{sec:planner}

The Planner receives the editing instruction and the original code. 
Its role is to generate a structured, variant-specific edit plan without producing any code. 
A separate edit plan is created for each visibility variant, and the Planner is restricted to using only the information available within that variant.

Each edit plan contains the following fields:
\begin{itemize}
    \item \textbf{Observed Elements}: Concrete code entities (function names, variable names, class definitions) extracted verbatim from the visible code.
    \item \textbf{Edit Intent}: A semantic description of the required transformation.
    \item \textbf{Target Description}: The location in the code where the edit should be applied.
    \item \textbf{Required Changes}: A list of specific modification actions (e.g., ``ADD parameter \texttt{timeout} to function \texttt{connect()}'').
    \item \textbf{Constraints}: Rules that must not be violated (e.g., ``preserve all existing function signatures'').
\end{itemize}

This separation between planning and code generation ensures that the Editor operates from grounded, unambiguous instructions rather than vague high-level directives.

\subsubsection{Editor Agent}
\label{sec:editor}

The Editor receives the structured edit plan along with the original code and produces the modified code. It is instructed to:

\begin{enumerate}
    \item Treat the \texttt{required\_changes} field as the single authoritative source of edits.
    \item Apply changes literally, not interpretively.
    \item Preserve original formatting, structure, and unrelated code.
    \item Produce only the edited code fragments, not full files.
\end{enumerate}

This protocol ensures minimal edits and promotes deterministic behavior, reducing the risk of “over-editing,” where unintended refactoring or cosmetic changes can break existing functionality.

\subsubsection{Verifier Agent}
\label{sec:verifier}

The Verifier executes the edited code against the full test suite in a secure sandbox environment. 
Unlike approaches that rely on an LLM to predict test outcomes, SAFEdit executes real unit tests, providing ground-truth pass/fail outcomes along with detailed error logs.

If all tests pass, the pipeline terminates successfully. 
If any test fails, the raw test logs are passed to the Failure Abstraction Layer (FAL) (Section~\ref{sec:fal}) for structured processing before being fed back into the refinement loop.

\subsection{Iterative Refinement Loop}
\label{sec:refinement}

Differentiating SAFEdit from single-shot approaches is its iterative refinement mechanism.
When the Verifier detects test failures, the pipeline enters a refinement cycle rather than terminating. The loop proceeds as follows:

\begin{enumerate}
    \item The Verifier's structured feedback (produced by the FAL) is fed back to the Editor.
    \item Based on the original edit plan and the failure feedback, the Editor generates a revised version of the code.
    \item The revised code is then executed against the test suite by the Verifier.
    \item This refinement cycle continues for up to three iterations.
\end{enumerate}

The maximum of three refinement iterations was selected to balance accuracy gains against computational cost.

\subsection{Failure Abstraction Layer (FAL)}
\label{sec:fal}

The FAL replaces raw test output with structured, actionable feedback for the Editor during refinement. 
Raw test logs are typically verbose, noisy, and may contain redundant stack traces that obscure the root cause of failures~\cite{pytest_docs_traceback}. 
The FAL transforms unstructured test logs into semantic failure reports through a three-stage process:

\subsubsection{Stage 1: Log Parsing}
In the first stage, the log parser systematically extracts structured information from raw test output, including:
\begin{itemize}
    \item The name of each failed test function
    \item The exception type (e.g., \texttt{AssertionError}, \texttt{NameError})
    \item Expected vs.\ actual values for assertion mismatches
\end{itemize}

\subsubsection{Stage 2: Failure Classifier}
In the second stage, the failure classifier maps each parsed failure to one of 14 structured failure types, including:
\texttt{SYNTAX\_ERROR}, \texttt{ASSERTION\_\allowbreak MISMATCH}, \texttt{IMPORT\_ERROR}, \texttt{ATTRIBUTE\_\allowbreak ERROR}, \texttt{TYPE\_ERROR} and \texttt{VALUE\_ERROR}

This classification is performed using deterministic pattern matching rules, ensuring no additional LLM cost and full reproducibility.

\subsubsection{Stage 3: Explanation Generator}
In the last stage, the explanation generator formats the classified failure into a structured feedback block with explicit diagnostic fields:

\begin{table}[h]
\small
\begin{tabular}{ll}
\texttt{[TEST]}       & \texttt{test\_calculator\_subtract} \\
\texttt{[TYPE]}       & \texttt{ASSERTION\_MISMATCH} \\
\texttt{[DIAGNOSIS]}  & Expected value does not match actual value. \\
\texttt{[ACTION]}     & Modify the logic to match the expected calculation. \\
\texttt{[EXPECTED]}   & 2 \\
\texttt{[ACTUAL]}     & 8 \\
\texttt{[CONFIDENCE]} & 95\% \\
\end{tabular}
\end{table}

This structured format provides the Editor with: (1)~what failed, (2)~why it failed, and (3)~a suggested repair action. The hypothesis is that this semantic abstraction reduces the number of iterations, prevents instruction hallucinations, and minimizes over-editing compared to raw logs.


%% file: sections/4_evaluation/evaluation.tex
\section{Evaluation}
\label{sec:evaluation}

In this section, we describe the experimental setup used to evaluate SAFEdit, including the benchmark dataset, baseline methods, evaluation metrics, and the automated error taxonomy used for qualitative failure analysis.

\subsection{Dataset}
\label{sec:dataset}

We evaluated SAFEdit on the EditBench benchmark~\cite{chi2025editbenchevaluatingllmabilities}, a large-scale dataset of real-world instructed code editing tasks. 
The full benchmark contains \textbf{540 tasks} spanning five natural languages (English, Polish, Spanish, Chinese, and Russian) and three programming language families (Python, JavaScript, and JavaScript/REACT).
Each task consists of: (1)~a natural-language editing instruction, (2)~the original source code, (3)~a comprehensive test suite defining correctness, (4)~an optional highlighted code region indicating the relevant function or block of code, and (5)~a cursor position marking the exact line of interest.

\subsubsection{Data Filtering and Selection}

To control for confounding variables related to multilingual instruction parsing and cross-language code generation, we removed records in which the three code snippets provided in each task (\texttt{CODE ONLY}, \texttt{HIGHLIGHT}, and \texttt{HIGHLIGHT + CURSOR\_POSITION}) are identical. 
Such cases do not constitute meaningful editing scenarios, as no actual modification is required.

After this step, 485 tasks remained. We then excluded 40 additional tasks due to malformed test suites, missing dependencies, or non-deterministic test behavior that could confound pass/fail signals. This process yielded a final evaluation set of 445 tasks, corresponding to 89 tasks for each of the five languages. All filtering criteria were applied uniformly across all evaluated methods.

\subsubsection{Visibility Variants}

Each task in the evaluation set was evaluated using three visibility variants (as defined in Section~\ref{sec:problem}): \texttt{CODE ONLY}, \texttt{HIGHLIGHT}, and \texttt{HIGHLIGHT + CURSOR\_POSITION}. 
This yielded $445 \times 3 = 1335$ individual evaluation instances, allowing us to assess how spatial context affected agent performance across all methods.
\subsection{Baselines}
\label{sec:baselines}

To evaluate the contribution of SAFEdit's design, we compared it to two baselines:

\begin{itemize}
    \item \textbf{Zero-Shot}: A single prompt-response interaction for instruction\allowbreak-guided code editing. The model receives the instruction and code, and produces the edited code in one pass, without feedback or iteration. For this baseline, we used the zero-shot results reported in the original EditBench study.

    \item \textbf{ReAct}~\cite{yao2023react}: A single-agent approach for instruction-guided code editing based on the reasoning--acting paradigm. The agent interleaves thought, action, and observation steps in a loop, using the same test runner for feedback. To ensure comparability, we implemented the ReAct baseline using the same underlying model (GPT-4.1) in the SAFEdit agentic system. Unlike SAFEdit, ReAct does not decompose the task into separate subtasks such as planning, editing, and verification.
\end{itemize}

For a fair comparison, in our evaluation, SAFEdit and both baselines operated under the same iteration budget (a maximum of three iterations) and used identical test execution infrastructure, ensuring that performance differences were attributable to architectural design rather than evaluation conditions.

\subsection{Automated Error Taxonomy}
\label{sec:taxonomy}

To perform a qualitative analysis of agent failures, we introduce an automated \emph{error taxonomy} that classifies why the agent fails, in addition to recording pass/fail outcomes. We define four root-cause categories for code editing failures:

\begin{enumerate}
    \item \textbf{Instruction Hallucination (IH)}: Instead of following the editing instruction, the agent misinterprets or ignores it; for example, instead of adding a log statement, the agent deletes the entire function.
    
    \item \textbf{Implementation Gap (IG)}: Instead of implementing the requested change correctly, the agent produces logically incorrect code; for example, implementing subtraction as addition.
    
    \item \textbf{Regression Error (RE)}: Instead of preserving existing functionality while satisfying the edit requirement, the agent introduces a new error; for example, adding a parameter without providing a default value.
    
    \item \textbf{Context Misalignment (CM)}: Instead of propagating the required change across all affected code regions, the agent updates only part of the relevant context; for example, renaming a function but not updating its call sites.
\end{enumerate}

\emph{Classification Pipeline. } For each failed task, the classifier receives the original instruction, the original source code, the agent's edited code, the test failure logs from the Verifier, and the Planner's edit plan. An LLM-based classifier (\texttt{GPT-4.1}) analyzes these inputs and assigns a category, along with a confidence score (0--1) and a textual justification. The taxonomy schema and classification prompt are defined using Pydantic models to ensure structured output that can be validated.

\subsection{Experimental Setup}
\label{sec:setup}

\subsubsection{Model Configuration}

All of SAFEdit's agents (Planner, Editor, and Verifier), as well as the baselines, use \textbf{GPT-4.1} as the backbone LLM. We set the temperature to 0.1 to encourage deterministic outputs while allowing minimal variation. The maximum refinement budget is fixed at \textbf{three iterations} for all methods.

\subsubsection{Orchestration}

SAFEdit's multi-agent pipeline is orchestrated using the CrewAI framework~\cite{barbarroxa2024benchmarking}, which manages agent sequencing, message passing, and tool invocations. Each agent operates with a dedicated system prompt and tool set, and agents communicate exclusively through structured intermediate outputs (edit plans and feedback reports) rather than free-form text.
We selected CrewAI, because its orchestration capabilities and modular design enabled us to implement a high-performing multi-agent framework while maintaining explicit role separation, deterministic task routing, and reproducible execution.

\subsubsection{Test Execution Environment}

All code edits were evaluated by running the associated test suite in a secure sandboxed environment. The sandbox provided process-level isolation, preventing side effects across tasks. Each task was executed under a 30-second timeout to guard against infinite loops and excessively resource-intensive behavior.
\subsection{Evaluation Metrics}
\label{sec:metrics}

We evaluate all methods using the following metrics:

\begin{itemize}
    \item \textbf{Task Success Rate (TSR)}: The primary metric. A task is considered successful if and only if all associated tests pass after the agent's final edit. The TSR is computed as $\mathrm{TSR} = \frac{\text{passed tasks}}{\text{total tasks}} \times 100\%$. We report the TSR both overall and per visibility variant.
    
    \item \textbf{Iteration Efficiency}: The average number of refinement iterations used across all tasks. Lower values indicate faster convergence.
    
    \item \textbf{First-Try Success Rate}: The proportion of tasks that pass on the first attempt, without any refinement. This metric reflects the quality of the initial edits.

    \item \textbf{Failure Category Distribution}: The distribution of failed tasks across the four error taxonomy categories defined in Section~\ref{sec:taxonomy}. This metric provides insight into the dominant failure modes of each method.
    
\end{itemize}

%% file: sections/5_results/results.tex
\section{Results}
\label{sec:results}

We evaluate \method{} on 445 EditBench tasks and compare its performance to that of single-agent baseline (ReAct) and strong single-model baselines reported in the original EditBench leaderboard~\cite{chi2025editbenchevaluatingllmabilities}. 
Unless stated otherwise, all reported numbers follow the benchmark protocol and use the TSR metric.
\subsection{Overall Performance}
\label{sec:results-overall}

\begin{figure}[!h]
    \centering
    \includegraphics[width=\columnwidth]{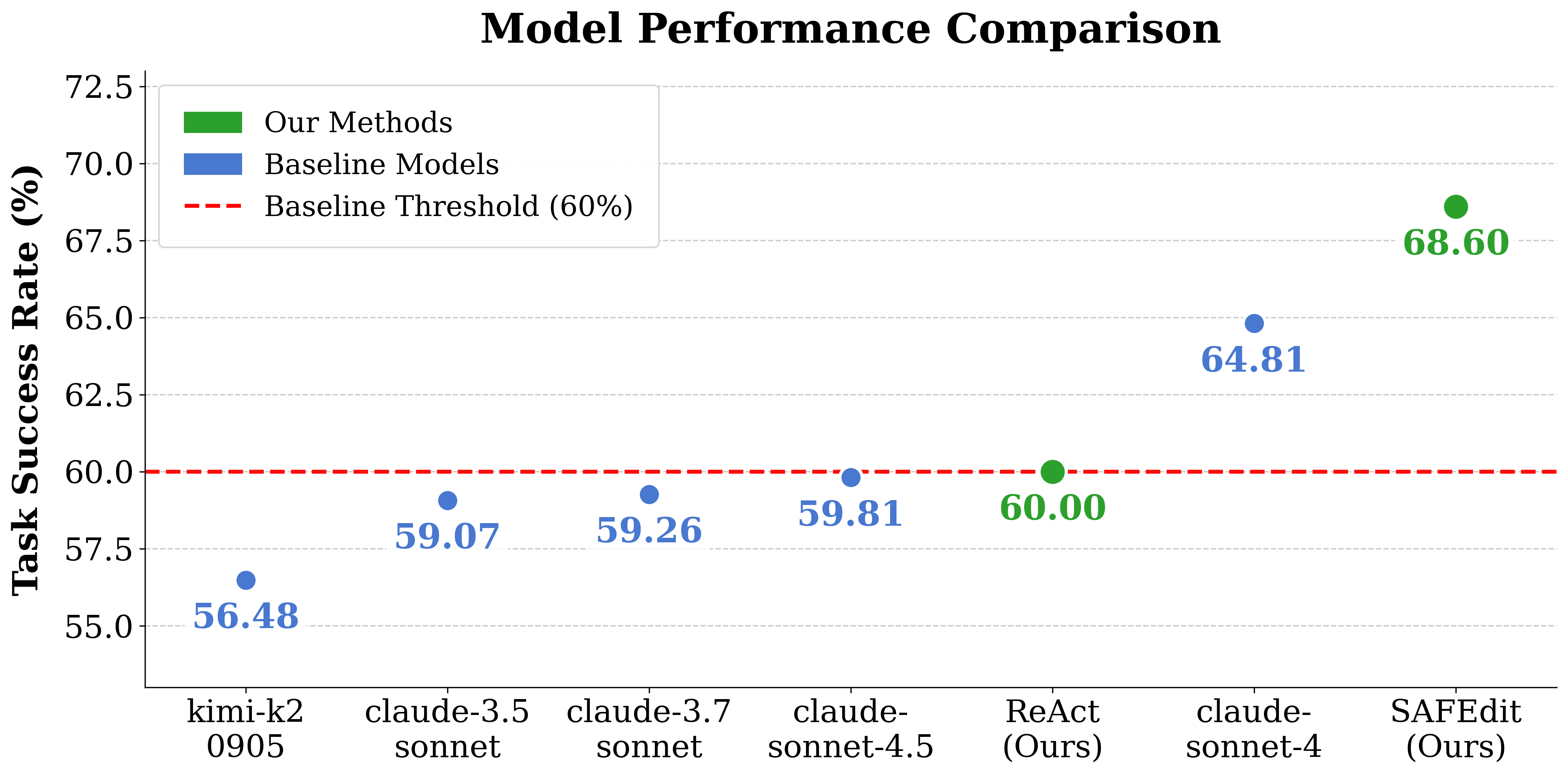}
    \caption{
    TSR under the \textsc{HIGHLIGHT} visibility variant for SAFEdit, ReAct baseline, and top-5 EditBench leaderboards baselines.
    }
    \label{fig:highlight_comparison}
\end{figure}

We begin with an analysis of the performance under the \textsc{HIGHLIGHT} visibility variants, which serves as the primary evaluation setting in EditBench. Figure~\ref{fig:highlight_comparison} presents the TSR obtained with SAFEdit, ReAct, and the top EditBench leaderboard baselines.

SAFEdit achieves \textbf{68.6\%}, the highest score of all the evaluated methods. 
This places SAFEdit \textbf{+3.8pp} above the strongest single-model baseline reported on the EditBench leaderboard (claude-sonnet-4, 64.8\%) and \textbf{+8.6pp} above ReAct (60.0\%), despite all baselines using the same backbone LLM configuration.

The 60\% threshold is particularly meaningful in this benchmark. In the original EditBench study, 39 of the 40 evaluated models failed to surpass this mark~\cite{chi2025editbenchevaluatingllmabilities}. While claude-sonnet-4 marginally exceeds it under the HIGHLIGHT variant, most leaderboard models remain below this line.

Three distinct performance tiers can be seen in the distribution presented in Figure~\ref{fig:highlight_comparison}:
(i) SAFEdit in the high-60s,
(ii) frontier single-model baselines and our ReAct baseline implementation clustered around 55–65\%, and
(iii) the remaining baselines from the EditBench leaderboard, which fall under 55\% and below the 60\% threshold.
This separation suggests that architectural decomposition likely contributes to the observed improvement beyond raw model scaling alone.

Overall, the HIGHLIGHT results demonstrate that SAFEdit’s structured Planner–Editor–Verifier design yields a substantial and measurable advantage over both single-agent (ReAct) and single-model baselines, confirming the effectiveness of multi-agent decomposition in reliable instructed code editing.

\subsection{Results Per Language}
\label{sec:results-languages}

Table~\ref{tab:tsr_languages} reports the TSR obtained for each instruction language under the HIGHLIGHT variant for the multi-agent decomposition and the ReAct baseline (EditBench does not report results separately for each language).

As can be seen, \method{} outperforms ReAct in each case, with improvements ranging from \textbf{+5.7pp} (Russian) up to \textbf{+12.4pp} (Spanish), and an average improvement of \textbf{+8.6pp}. This consistency suggests that the observed benefit is not language specific and is consistent with the role-separated design of planning, editing, and verification.

\begin{table}[!h]
\centering
\caption{TSR (\%) per language for multi-agent decomposition and ReAct baseline (EN - English, PL - Polish, ES - Spanish, CH - Chinese, RU - Russian).}
\label{tab:tsr_languages}
\small
\begin{tabular}{lcccccc}
\toprule
\textbf{Method} & \textbf{EN} & \textbf{PL} & \textbf{ES} & \textbf{CH} & \textbf{RU} & \textbf{Avg} \\
\midrule
ReAct & 61.8 & 57.3 & 57.3 & 60.7 & 62.9 & \textbf{60.0} \\
\rowcolor{green!15}
\textbf{SAFEdit} & \textbf{70.5} & \textbf{67.0} & \textbf{69.7} & \textbf{67.0} & \textbf{68.6} & \textbf{68.6} \\
\midrule
\textbf{$\Delta$ - Delta} & \textcolor{green!60!black}{\textbf{+8.7}} & \textcolor{green!60!black}{\textbf{+9.7}} & \textcolor{green!60!black}{\textbf{+12.4}} & \textcolor{green!60!black}{\textbf{+9.3}} & \textcolor{green!60!black}{\textbf{+5.7}} & \textcolor{green!60!black}{\textbf{+8.6}} \\
\bottomrule
\end{tabular}
\normalsize
\end{table}

The absolute TSR varies across languages for both methods. While the TSR of \method{} remains strong overall (67--71\% range), ReAct's TSR drops more sharply in some non-English languages, particularly in Chinese and Polish. This pattern aligns with the intuition that multilingual instructed editing amplifies ambiguity and increases reliance on robust task interpretation. A structured planning stage likely helps preserve intent before editing is performed.

\subsection{Variant Analysis Per Language}
\label{sec:results-variants}

To better understand context sensitivity, Table~\ref{tab:tsr_variants} breaks down the TSR by visibility variant for each language for multi-agent decomposition and ReAct baseline. Two trends emerge. First, for English, \method{} is fully consistent across variants (70.5\% in all three variants), indicating that the Planner can reliably infer edit location and intent from the instruction and code context without depending on cursor grounding. Second, for several non-English languages (e.g., Polish and Spanish), adding cursor information occasionally introduces small regressions, suggesting that spatial signals can become noisy when instruction understanding is weaker or when the highlighted region already provides sufficient anchoring.

\begin{table}[!h]
\centering
\caption{TSR (\%) per visibility variant and language. The delta values (red/grey) are relative to the CODE ONLY variant.}
\label{tab:tsr_variants}
\small
\setlength{\tabcolsep}{2pt}
\begin{tabular}{llccc}
\toprule
\textbf{Method} & \textbf{Lang.} & \textbf{CODE ONLY} & \textbf{HIGHLIGHT} & \textbf{HIGHLIGHT + CURSOR} \\
\midrule
\rowcolor{green!15}
SAFEdit & EN & 70.5 & 70.5 \textcolor{gray}{\small(0.0)} & 70.5 \textcolor{gray}{\small(0.0)} \\
\rowcolor{green!15}
SAFEdit & PL & 69.3 & 67.0 \textcolor{red}{\small($-$2.3)} & 66.3 \textcolor{red}{\small($-$3.0)} \\
\rowcolor{green!15}
SAFEdit & ES & 69.7 & 69.7 \textcolor{gray}{\small(0.0)} & 68.5 \textcolor{red}{\small($-$1.2)} \\
\rowcolor{green!15}
SAFEdit & CH & 67.0 & 67.0 \textcolor{gray}{\small(0.0)} & 66.3 \textcolor{red}{\small($-$0.7)} \\
\rowcolor{green!15}
SAFEdit & RU & 72.1 & 68.6 \textcolor{red}{\small($-$3.5)} & 68.6 \textcolor{gray}{\small(0.0)} \\
\midrule
ReAct   & EN & 64.0 & 61.8 \textcolor{red}{\small($-$2.2)} & 61.8 \textcolor{red}{\small($-$2.2)} \\
ReAct   & PL & 57.3 & 57.3 \textcolor{gray}{\small(0.0)} & 57.3 \textcolor{gray}{\small(0.0)} \\
ReAct   & ES & 57.3 & 57.3 \textcolor{gray}{\small(0.0)} & 57.3 \textcolor{gray}{\small(0.0)} \\
ReAct   & CH & 61.8 & 60.7 \textcolor{red}{\small($-$1.1)} & 60.7 \textcolor{red}{\small($-$1.1)} \\
ReAct   & RU & 62.9 & 62.9 \textcolor{gray}{\small(0.0)} & 62.9 \textcolor{gray}{\small(0.0)} \\ 
\bottomrule
\end{tabular}
\normalsize
\end{table}

ReAct shows a different profile: performance is largely flat across the variants in most non-English languages, implying limited ability to systematically exploit the additional spatial signal. In English, however, the TSR of ReAct drops by \textbf{-2.2pp} when moving from \textsc{CODE ONLY} to \textsc{HIGHLIGHT}, and does not change with cursor information. Overall, these results reveal cursor grounding is not universally beneficial and can be counterproductive in some multilingual settings.
\subsection{Iterative Refinement Contribution}
\label{sec:results-refinement}

Table~\ref{tab:iterations} quantifies the contribution of the refinement loop. Across the evaluated languages, \method{} passes about half of the tasks tests on the first iteration (47--54\%), and reaches its reported TSR through iterative correction, yielding gains of \textbf{+14.2pp} to \textbf{+22.8pp} depending on the language. This confirms that a single-pass edit is often insufficient, even with SAFEdit's strong Planner, and that verification-guided refinement is a major driver of the performance.

\begin{table}[!h]
\centering
\caption{Iteration efficiency TSR values for SAFEdit. First-Try = proportion of tasks passing their tests without any refinement.}
\label{tab:iterations}
\small
\begin{tabular}{lccc}
\toprule
\textbf{Language} & \textbf{First-Try (\%)} & \textbf{Avg Iterations} & \textbf{Final TSR (\%)} \\
\midrule
English (EN) & 47.7  & 2.02 & 70.5  \textcolor{green!60!black}{\small(+22.8)} \\
Polish (PL)  & 52.3  & 1.86 & 67.0 \textcolor{green!60!black}{\small(+14.7)} \\
Spanish (ES) & 49.4  & 1.92 & 69.7 \textcolor{green!60!black}{\small(+20.3)} \\
Chinese (CH) & 52.8  & 1.83 & 67.0 \textcolor{green!60!black}{\small(+14.2)} \\
Russian (RU) & 53.9  & 1.81 & 68.6 \textcolor{green!60!black}{\small(+14.7)} \\
\midrule
\textbf{Total Average} & 51.2 & 1.8 & 68.6 \textcolor{green!60!black}{\small(+17.4)} \\
\bottomrule
\end{tabular}
\normalsize
\end{table}

Importantly, convergence is efficient: the average number of iterations required is close to two (out of a maximum of three), indicating that most failures are resolved in one additional correction step. Qualitatively, this behavior aligns with the intended design of \method{}: the Verifier produces targeted feedback, enabling the Editor to apply localized repairs rather than re-derive the entire edit from scratch.

\subsection{Error Taxonomy}
\label{sec:results-taxonomy}
\begin{figure*}[t]
    \centering
    \includegraphics[width=\linewidth, ,height=10cm, keepaspectratio]{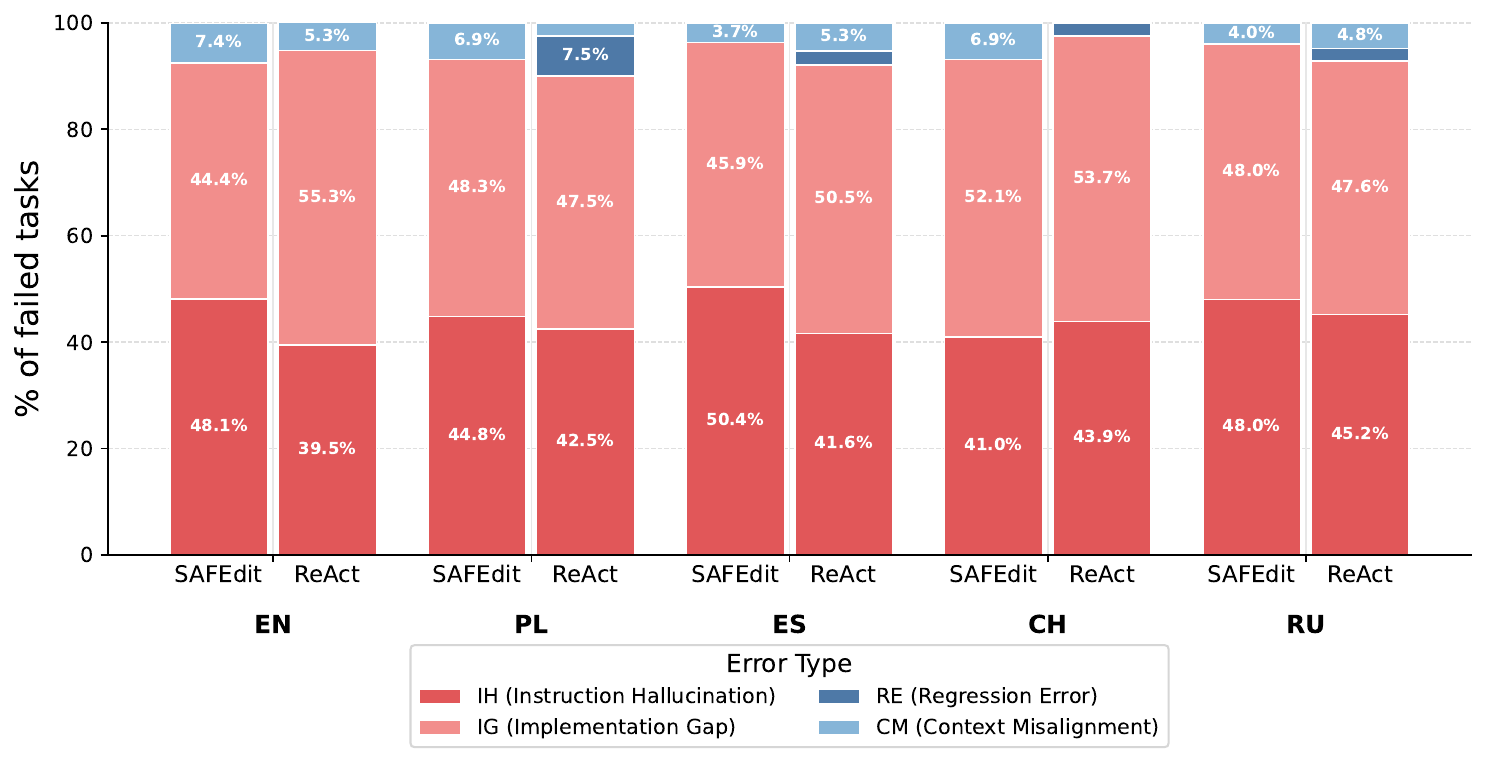}
    \caption{Stacked error distribution (\%) for failed tasks tests across the evaluated languages. For each language, SAFEdit multi-agent decomposition (left) and ReAct implementation (right) are broken down into the root-cause categories of IH, IG, RE, and CM. The figure does not include the EditBench models because it does not report results for each language.}
    \label{fig:error_taxonomy}
\end{figure*}

Figure~\ref{fig:error_taxonomy} presents the results of the error taxonomy analysis we conducted for the multi-agent decomposition and the ReAct baseline.
Across the evaluated languages, both methods fail primarily due to instruction hallucination (IH) and implementation gap (IG), but their failure profiles differ. ReAct's failures are IG-dominant in most languages (e.g., English (EN): 55.3\%, Spanish (ES): 50.5\%, Chinese (CH): 53.7\%), suggesting that the model often identifies the intended change but fails to implement it correctly or completely. This pattern is more consistent with execution/implementation limitations and insufficient effective context utilization (rather than a strict ``context window'' limitation), where relevant constraints from the codebase are not consistently integrated in the produced patch.

In contrast, SAFEdit exhibits a more balanced split between the root-cause categories of IH and IG in several languages (English, Polish, Russian), pointing to two distinct residual challenges: (1) correctly grounding the requested edit in the instruction (IH), and (2) translating the intended plan into a correct patch (IG). Notably, SAFEdit exhibits language-dependent failure modes: some languages fail primarily on IH (English and Spanish), whereas others fail primarily on IG (Polish and Chinese). Finally, a key reliability signal is that SAFEdit incurs \textbf{no Regression Errors (RE)} in any language (0.0\%), whereas ReAct exhibits non-zero regression errors in most languages (e.g., Polish, Spanish, Chinese and Russian). This suggests that the multi-agent workflow, together with verification-guided refinement, reduces unintended regressions in previously correct behavior and improves the localization of changes.

%% file: sections/6_discussion/discussion.tex
\section{Discussion}
Our study examined whether structured multi-agent decomposition can improve reliability in instructed code editing, a setting that requires precise modification of existing code under executable correctness constraints. Across the evaluated languages and visibility variants, SAFEdit consistently outperformed the compared baselines, suggesting that success in this task depends not only on model capability but also on the structure of the editing process and the use of execution-grounded refinement. Taken together, these findings support the view that instructed code editing is not merely a smaller version of code generation, but a distinct problem that benefits from explicit planning, constrained execution, and test-grounded verification.

\subsection{Structured Agent Decomposition Improves Editing Reliability}
\label{sec:disc_architecture}

A central finding of this study is that SAFEdit's structured decomposition into Planner, Editor, and Verifier agents yielded a clear performance advantage over unstructured single-model approaches. Although SAFEdit and ReAct were instantiated with the same underlying model, SAFEdit achieved a consistently higher TSR across all five languages, suggesting that the gains are consistent with benefits from decomposing the editing process rather than from backbone differences alone.

This advantage is consistent with the separation of roles between agents. ReAct operates as a single agent that interleaves reasoning and action within one loop, requiring the same process to interpret the instruction, generate the code change, and react to execution feedback. In contrast, SAFEdit distributes these responsibilities across specialized roles. The Planner defines the intended modification, the Editor performs the code change, and the Verifier provides execution-grounded feedback. This decomposition not only reduces the burden on any single reasoning step, but also allows each agent to specialize in a narrower subtask, resulting in a more controlled and reliable refinement process.

The results further suggest that iterative refinement, supported by structured verification, is not merely a recovery mechanism but a key contributor to the achieved performance. Rather than generating entirely new edits after each failed attempt, SAFEdit refines its behavior through structured diagnostic feedback derived from the FAL. This allows subsequent edits to target concrete failure causes more precisely. The fact that SAFEdit typically converged in less than two refinement iterations, while still obtaining large gains over first-pass performance, indicates that the interaction between decomposition and feedback was both effective and efficient.

\subsection{Robustness to Spatial Context and the Context-Exposure Paradox}
\label{sec:disc_context}

Another important observation is that SAFEdit's performance remained stable across the three visibility variants, whereas single-model baselines were more sensitive to how spatial context was presented. This finding is particularly notable, because one might expect additional localization cues, such as highlighted regions or cursor position, to make the editing task easier. Instead, both the original EditBench and our ReAct baseline results suggest that increasing spatial guidance can sometimes degrade performance in single-agent systems.

SAFEdit did not exhibit this pattern. In English, it maintained identical peak performance across all three variants, and in the non-English settings it showed only minor fluctuations. This supports the understanding that SAFEdit is not simply benefiting from greater context but instead is more robust to variation in how context is presented. In other words, structured agent decomposition appears to reduce sensitivity to prompt formatting and auxiliary spatial signals.

A plausible explanation is that the Planner already extracts and organizes the relevant editing intent before code modification begins. Once the task is translated into a specific edit plan, the Editor can operate on a more stable representation of the required change, rather than relying directly on raw prompt cues such as cursor markers. This may explain why SAFEdit remained stable across the English variants and showed only minor fluctuations in the non-English settings when additional context signals were introduced. From a practical perspective, such robustness is valuable, because real-world developer interactions are unlikely to provide perfectly standardized spatial guidance.

\subsection{Failure Analysis Reveals Qualitative Differences, Not Only Quantitative Ones}
\label{sec:disc_failures}

The automated error taxonomy provides an additional perspective on system behavior that is not captured by TSR alone. While overall success rates show that SAFEdit performs better, the failure distributions help explain how its reasoning differs from that of the ReAct baseline.

ReAct's failures were dominated by IG errors across most languages, suggesting that the agent often recognized the intended direction of the change but failed to implement it accurately in code. In contrast, SAFEdit showed a more balanced distribution across the IH and IG taxonomy categories. This pattern is consistent with its staged architecture: failures may arise either because the Planner did not fully capture the instruction or because the Editor did not accurately implement an otherwise reasonable plan.

One particularly noteworthy result is that SAFEdit produced no RE in any of the five languages. In contrast, ReAct occasionally introduced edits that satisfied one requirement while breaking previously correct behavior. In realistic development settings, avoiding collateral damage is often as important as implementing the requested change itself. SAFEdit's zero-regression profile suggests that structured planning, constrained editing, and verification feedback together help preserve existing functionality when applying targeted modifications.

More broadly, these findings show why pass/fail evaluation alone is insufficient for agentic editing systems. Two systems may fail at similar rates in some settings yet differ substantially in the nature of those failures. By highlighting these distinctions, the taxonomy strengthens the interpretability of the evaluation and helps identify where future structure improvements are needed.

\subsection{Implications for Trustworthy Code Editing}
\label{sec:disc_implications}

Collectively, the results suggest that reliable code editing requires more than stronger general-purpose language models. The task depends on reliably interpreting an instruction, localizing the intended modification, preserving surrounding invariants, and validating the outcome under execution. SAFEdit addresses these requirements not through a larger backbone, but through process structure: structured planning, bounded editing, and test-based refinement.

This has two broader implications. First, it suggests that benchmark advances in instructed code editing should not be interpreted solely as a property of model scale or raw generative ability. Architectural choices can influence performance and robustness, even when the same underlying model is used. Second, it highlights the value of evaluation designs  that consider  the final success rate as well as visibility-sensitive testing and structured failure analysis. Such evaluation is particularly important for agentic systems, where intermediate reasoning and coordination may shape outcomes in ways that aggregate metrics alone do not reflect.

\subsection{Computational Cost and Latency}
\label{sec:cost}
SAFEdit's design includes two cost-favorable properties. The FAL introduces no additional LLM cost due to its deterministic pattern-matching classifier, and the three-iteration budget is rarely fully consumed, with an average of 1.8 iterations indicating that most failures resolve in a single correction step.

Nonetheless, SAFEdit's performance gains come with increased computational cost relative to zero-shot single-model approaches, requiring at minimum three sequential LLM calls per task, rising to approximately five to six when iterative refinement is included. This overhead may be a practical consideration in latency-sensitive or high-volume settings.

Overall, the cost increase appears modest relative to the reliability benefits gained, particularly when compared to the more appropriate ReAct baseline 
where SAFEdit's primary structural overhead relative to ReAct baseline is the Planner's upfront call, with remaining costs comparable across both architectures.

\subsection{Limitations and Future Work}
\label{sec:threats}

This study has several limitations. First, although EditBench provides realistic and test-validated editing instances, our final evaluation set was curated after excluding malformed or unstable samples. This improves experimental control, but may reduce the diversity of evaluation cases relative to the full benchmark. Second, all of the methods were evaluated under the same bounded iteration budget and with the same backbone family, which supports fair comparison but does not address how SAFEdit would behave under other models or longer refinement horizons.

Future work should evaluate structured agentic editing on larger and more diverse repositories, compare multiple backbone models within the same framework, and examine longer-horizon editing scenarios that require broader propagation across files and dependencies. Additional work is also needed to strengthen failure analysis, including human validation of taxonomy labels and richer evaluation signals beyond binary test outcomes.

Overall, our findings show that structured role division and test-grounded feedback contribute to more reliable and trustworthy instructed code editing. They also suggest that progress in this setting may depend as much on system design as on model capability alone.

%% file: sections/7_conclusion/conclusion.tex
\section{Conclusion}

In this paper, we studied whether structured multi-agent decomposition can improve the reliability of instructed code editing, a task that requires precise modifications under executable correctness constraints. We introduced SAFEdit, a framework that separates the editing process into planning, editing, and verification stages, combined with execution-based iterative refinement.

Our results on the EditBench benchmark show that SAFEdit consistently outperforms strong single-agent and single-model baselines across multiple languages and context settings, achieving higher task success rates while reducing regression errors. These findings suggest that reliability in instructed code editing depends not only on model capability, but also on the structure of the editing process and the integration of test-grounded feedback.

Overall, our findings highlight the importance of structured reasoning, constrained editing, and execution-based validation for building more reliable code-editing systems.

%% file: references.bib
@misc{jiang2025ledextrainingllmsbetter,
      title={{LeDex}: Training {LLMs} to Better Self-Debug and Explain Code}, 
      author={Nan Jiang and Xiaopeng Li and Shiqi Wang and Qiang Zhou and Soneya Binta Hossain and Baishakhi Ray and Varun Kumar and Xiaofei Ma and Anoop Deoras},
      year={2025},
      eprint={2405.18649},
      archivePrefix={arXiv},
      primaryClass={cs.CL},
      url={https://arxiv.org/abs/2405.18649}, 
}

@misc{prasad2025learninggenerateunittests,
      title={Learning to Generate Unit Tests for Automated Debugging}, 
      author={Archiki Prasad and Elias Stengel-Eskin and Justin Chih-Yao Chen and Zaid Khan and Mohit Bansal},
      year={2025},
      eprint={2502.01619},
      archivePrefix={arXiv},
      primaryClass={cs.SE},
      url={https://arxiv.org/abs/2502.01619}, 
}

@inproceedings{
chi2025editbenchevaluatingllmabilities,
title={EditBench: Evaluating {LLM} Abilities to Perform Real-World Instructed Code Edits},
author={Wayne Chi and Valerie Chen and Ryan Shar and Aditya Mittal and Jenny Liang and Wei-Lin Chiang and Anastasios Nikolas Angelopoulos and Ion Stoica and Graham Neubig and Ameet Talwalkar and Chris Donahue},
booktitle={The Fourteenth International Conference on Learning Representations},
year={2026},
url={https://openreview.net/forum?id=FtL9eEmU6v}
}

@misc{cassano2024editevaluatingabilitylarge,
      title={Can It Edit? Evaluating the Ability of Large Language Models to Follow Code Editing Instructions}, 
      author={Federico Cassano and Luisa Li and Akul Sethi and Noah Shinn and Abby Brennan-Jones and Jacob Ginesin and Edward Berman and George Chakhnashvili and Anton Lozhkov and Carolyn Jane Anderson and Arjun Guha},
      year={2024},
      eprint={2312.12450},
      archivePrefix={arXiv},
      primaryClass={cs.SE},
      url={https://arxiv.org/abs/2312.12450}, 
}

@inproceedings{barbarroxa2024benchmarking,
  title={Benchmarking large language models for multi-agent systems: A comparative analysis of {AutoGen}, {CrewAI}, and {TaskWeaver}},
  author={Barbarroxa, Rafael and Gomes, Luis and Vale, Zita},
  booktitle={International Conference on Practical Applications of Agents and Multi-Agent Systems},
  pages={39--48},
  year={2024},
  organization={Springer},
  url = {https://doi.org/10.1007/978-3-031-70415-4_4},

}

@inproceedings{yao2023react,
  title={{ReAct}: Synergizing Reasoning and Acting in Language Models},
  author={Yao, Shunyu and Zhao, Jeffrey and Yu, Dian and Du, Nan and Shafran, Izhak and Narasimhan, Karthik and Cao, Yuan},
  booktitle={The Eleventh International Conference on Learning Representations (ICLR)},
  year={2023},
  url={https://arxiv.org/abs/2210.03629}
}

@misc{gu2025llmtestgenerationiterative,
      title={LLM Test Generation via Iterative Hybrid Program Analysis}, 
      author={Sijia Gu and Noor Nashid and Ali Mesbah},
      year={2025},
      eprint={2503.13580},
      archivePrefix={arXiv},
      primaryClass={cs.SE},
      doi={https://doi.org/10.1145/3744916.3764553},
      url={https://arxiv.org/abs/2503.13580}, 
}

@misc{chen2021evaluatinglargelanguagemodels,
      title={Evaluating Large Language Models Trained on Code}, 
      author={Mark Chen and Jerry Tworek and Heewoo Jun and Qiming Yuan and Henrique Ponde de Oliveira Pinto and Jared Kaplan and Harri Edwards and Yuri Burda and Nicholas Joseph and Greg Brockman and Alex Ray and Raul Puri and Gretchen Krueger and Michael Petrov and Heidy Khlaaf and Girish Sastry and Pamela Mishkin and Brooke Chan and Scott Gray and Nick Ryder and Mikhail Pavlov and Alethea Power and Lukasz Kaiser and Mohammad Bavarian and Clemens Winter and Philippe Tillet and Felipe Petroski Such and Dave Cummings and Matthias Plappert and Fotios Chantzis and Elizabeth Barnes and Ariel Herbert-Voss and William Hebgen Guss and Alex Nichol and Alex Paino and Nikolas Tezak and Jie Tang and Igor Babuschkin and Suchir Balaji and Shantanu Jain and William Saunders and Christopher Hesse and Andrew N. Carr and Jan Leike and Josh Achiam and Vedant Misra and Evan Morikawa and Alec Radford and Matthew Knight and Miles Brundage and Mira Murati and Katie Mayer and Peter Welinder and Bob McGrew and Dario Amodei and Sam McCandlish and Ilya Sutskever and Wojciech Zaremba},
      year={2021},
      eprint={2107.03374},
      archivePrefix={arXiv},
      primaryClass={cs.LG},
      url={https://arxiv.org/abs/2107.03374}, 
}

@misc{austin2021programsynthesislargelanguage,
      title={Program Synthesis with Large Language Models}, 
      author={Jacob Austin and Augustus Odena and Maxwell Nye and Maarten Bosma and Henryk Michalewski and David Dohan and Ellen Jiang and Carrie Cai and Michael Terry and Quoc Le and Charles Sutton},
      year={2021},
      eprint={2108.07732},
      archivePrefix={arXiv},
      primaryClass={cs.PL},
      url={https://arxiv.org/abs/2108.07732}, 
}

@misc{jimenez2024swebenchlanguagemodelsresolve,
      title={SWE-bench: Can Language Models Resolve Real-World GitHub Issues?}, 
      author={Carlos E. Jimenez and John Yang and Alexander Wettig and Shunyu Yao and Kexin Pei and Ofir Press and Karthik Narasimhan},
      year={2024},
      eprint={2310.06770},
      archivePrefix={arXiv},
      primaryClass={cs.CL},
      url={https://arxiv.org/abs/2310.06770}, 
}

@article{guo2024codeeditorbench,
  author = {Jiawei Guo and others},
  title = {CodeEditorBench: Evaluating Code Editing Capability of Large Language Models},
  journal = {arXiv preprint arXiv:2404.03543},
  year = {2024},
  url = {https://arxiv.org/abs/2404.03543},
  note = {[cs.SE]}
}

@inproceedings{ruiz2025art,
  author = {Javier Vallecillos Ruiz and others},
  title = {The Art of Repair: Optimizing Iterative Program Repair with Instruction-Tuned Models},
  booktitle = {International Conference on Evaluation and Assessment in Software Engineering (EASE)},
  year = {2025},
  url  = {https://arxiv.org/abs/2505.02931},
}

@misc{huang2024agentcodermultiagentbasedcodegeneration,
      title={AgentCoder: Multi-Agent-based Code Generation with Iterative Testing and Optimisation}, 
      author={Dong Huang and Jie M. Zhang and Michael Luck and Qingwen Bu and Yuhao Qing and Heming Cui},
      year={2024},
      eprint={2312.13010},
      archivePrefix={arXiv},
      primaryClass={cs.CL},
      url={https://arxiv.org/abs/2312.13010}, 
}

@article{chen2024coder,
  author = {Zhibo Chen and others},
  title = {CodeR: Issue Resolving with Multi-Agent and Task Graphs},
  journal = {arXiv preprint arXiv:2406.01304},
  year = {2024},
  url = {https://arxiv.org/abs/2406.01304}
}

@manual{pytest_docs_traceback,
  title        = {pytest Traceback Styles},
  author       = {{pytest development team}},
  organization = {pytest},
  year         = {2024},
  note         = {Available at: https://docs.pytest.org/en/stable/how-to/output.html\#modifying-python-traceback-printing},
}
